# *In-situ* molecular-level observation of methanol catalysis at the water-graphite interface

*William Foster[§], Juan A. Aguilar[†], Halim Kusumaatmaja[§] and Kislon Voïtchovsky[§]\**

[§]Durham University, Physics Department, Durham DH1 3LE
[†]Durham University, Chemistry Department, Durham DH1 3LE

**Abstract**

Methanol occupies a central role in chemical synthesis and is considered an ideal candidate for cleaner fuel storage and transportation. It can be catalyzed from water and volatile organic compounds such as carbon dioxide, thereby offering an attractive solution for reducing carbon emissions. However molecular-level experimental observations of the catalytic process are scarce, and most existing catalysts tend to rely on empirically optimized, expensive and complex nano-composite materials. This lack of molecular-level insights has precluded the development of simpler, more cost-effective alternatives. Here we show that graphite immersed in ultrapure water is able to spontaneously catalyze methanol from volatile organic compounds in ambient conditions. Using single-molecule resolution atomic force microscopy (AFM) in liquid, we directly observe the formation and evolution of methanol-water nanostructures at the surface of graphite. These molecularly ordered structures nucleate near catalytically active surface features such as atomic step edges and grow progressively as further methanol is being catalyzed. Complementary nuclear magnetic resonance analysis of the liquid confirms the formation of methanol and quantifies its concentration. We also show that electric fields significantly enhance the catalysis rate, even when as small as that induced by the natural surface potential of the silicon AFM tip. These findings could have a significant impact on the development of organic catalysts and on the function of nanoscale carbon devices.

**Introduction**

The conversion of unwanted volatile organics such as carbon dioxide to methanol is of high interest given the pressing need for alternative energy sources to fossil fuel[1] and the significant potential to reduce carbon emissions.[2] Methanol also functions as an important platform molecule for chemical synthesis and offers an ideal solution for cleaner energy storage and transportation.[3] At the present time, methanol is catalyzed on an industrial scale,[4] but usually at high temperatures and pressures and relying on catalysts made of complex composite materials typically comprising active metal nanoparticles in an oxide support .[5–7] Given the complexity of these composites, their catalytic behavior is still not fully understood although the synergy between the constituent components has been shown to be one of the key elements.[8] Significantly, composites tend to require a complex nanoscale arrangement, making them expensive and highly sensitive to even slight structural changes. There is hence a strong need for simpler and cheaper alternatives that can be easily sourced and replaced.

Organic materials such as graphite derivatives present obvious candidates as alternative catalysts.[9] The use of graphitic materials in science and technology has grown dramatically over the last decade owing to graphene's extraordinary electronic and physical properties. Current applications range from electrochemical devices and fuel cells[10] to energy storage,[11] photovoltaics[12] and the development of materials exhibiting unique mechanical properties.[13] Pure graphene is hydrophobic and the ambient humidity can influence the behavior of devices[14,15] due to the formation of a nanoscopic water layer on all exposed surfaces. The hydrophilic derivative of graphene, graphene oxide (GrO), is obtained by the replacement of single carbon atoms in graphene sheets with oxygen containing functional groups such as epoxy and methoxy groups. GrO retains some of the unique electronic properties of graphene, but the presence of hydrophilic groups makes it soluble in water, opening new avenues for applications in water filtration and ion sieving[16] along with molecular sensing.[17] Recent results suggest that GrO can act as a photocatalyst for the conversion of water and carbon dioxide to methanol.[18] This behavior is attributed to the presence of the hydrophilic functional groups that stretch the bandgap of GrO, hence allowing the photo-generated electrons and holes to serve as oxidation and reduction sites. Such catalytic effects have never been observed for pure graphene where the regular bandgap and the absence of chemical singularities do not favor localized electrons.

The surface of bulk graphite presents singularities at exposed atomic steps and edges. These singularities have long been known to make the edges of graphite electrochemically active.[19] Recent studies have also demonstrated that graphite's basal plane, previously considered electrochemically inert, has an activity comparable to that of noble metal electrodes such as platinum.[20] These findings suggest that graphite may offer a suitable alternative to metal electrodes given the fact that it can be readily immersed in aqueous solutions, unlike graphene.

**Results and Discussion**

Here we demonstrate catalytic production of methanol at the surface of immersed highly orientated pyrolytic graphite (HOPG) in ambient conditions. The process occurs spontaneously with the thermal energy available, but is stimulated in the presence of an applied electric field. We quantify the amount of methanol produced using $^1$H nuclear magnetic resonance (NMR) spectroscopy of the resulting solution and follow *in-situ* the process with single-molecule resolution atomic force microscopy (AFM) in liquid. This includes the resulting self-assembly of water and alcohol molecules at the HOPG-water interface. Representative examples of this self-assembly can be seen in Figure 1a, where a molecularly structured patch is slowly growing in a system that initially consisted only of ultrapure water at the surface of HOPG. Consistent with a catalytic reaction, the nucleating structures are seen predominantly at the more electrochemically active edges of HOPG and are first observed after scanning for some time, typically more than an hour. Once stable, the molecular details of the growing domains can be imaged with AFM, often revealing assemblies involving units comparable to the size of water and short alcohol molecules (Figure 1b). Pure liquid water itself cannot form long-lived structures on HOPG at room temperature,[21] indicating that the observed patch must contain molecules formed *in-situ*.

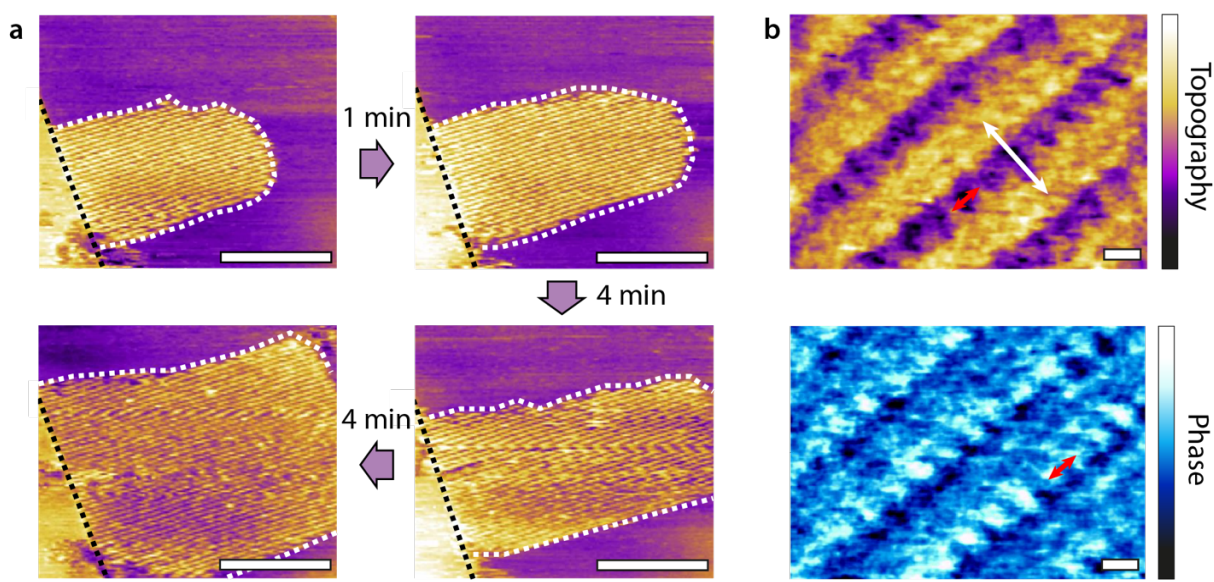

**Figure 1**. High resolution amplitude modulation AFM imaging of HOPG immersed in initially ultrapure water. (a) A solid-like patch formed by the self-assembly of molecules (dashed white outline) nucleates from an atomic step at the HOPG surface (dashed black line). The molecular self-assembly is observed here *in-situ* as it progressively grows across the HOPG surface over a period of 9 minutes, with the patch edges moving away from the step. Row-like structures with a periodicity of 4.30 ± 0.28 nm as visible within the patch. (b) Sub-nanometer imaging of other structures reveals detailed features (0.79 ± 0.08 nm periodicity, red arrows) perpendicular to the main rows (periodicity 2.45 ± 0.08 nm, white arrow). The exact molecular arrangement is not known, but strongly reminiscent of the alternated water-methanol nano-ribbons recently reported by our group.[22] The white scale bars are 100 nm in (a) and 1nm in (b). The purple color scale bar represents a topographic variation of 20 Å in (a) and 1 Å nm in (b). The blue

scale bar represents a phase variation of 20º in (a) and 10º in (b). In (a) the time lapse between the first and second frames is 1 minute and then 4 minutes elapses between the subsequent frames.

A recent study by our group has shown that water and methanol can self-assemble into stable, layered structures[22] at the interface with HOPG. Combining high-resolution AFM with computational simulations, these structures were shown to be stabilized by a group effect, through an extended network of hydrogen bonds involving both water and alcohol molecules. The relatively weak interaction between the solvent and the graphite surface enabled a high degree of polymorphism in the molecular arrangement of the self-assembled structures. The formation of row-like features with periodicities up to 6 nm was the most commonly observed supramolecular assembly. The formation of the patterns did not appear to be influenced by the scanning AFM tip, but was instead driven by the underlying graphite with the rows orientation in registry with the graphite lattice. Here, the molecular assemblies developing at the interface with HOPG in ultrapure water (Figure 1) are consistent with those reported for methanol-water mixtures[22] suggesting that HOPG-induced catalysis of water into methanol is occurring.

To independently confirm the formation of methanol, we conducted NMR analysis on the solution in direct contact with the HOPG before and after nucleation was observed. Practically, the measurement is challenging because it requires observing small quantities of methanol (typically sub-millimolar) in the presence of a signal (water) that is five orders of magnitude larger. Therefore, it was necessary to suppress the water signal very efficiently to allow for unambiguous identification of the methanol produced. This was achieved with the recently reported Robust-5 pulse sequence[23] (see experimental section for details). The results, presented in Figure 2, compare three sets of measurements: (i) ultrapure water placed for 5 seconds in contact with the surface of HOPG, (ii) ultrapure water placed for 2 hours in contact with the surface of HOPG, and (iii) ultrapure water placed for 2 hours in contact with the surface of HOPG while applying a DC potential of +1 V to the HOPG surface with respect to a platinum electrode placed directly in the water. The value of the applied voltage was selected to avoid any chemical modification of the surface.[24] All the samples were collected in identical conditions, at room temperature, in contact with air and over HOPG previously heated above 120 ºC to evaporate any historical contaminant (see experimental section).[25] Methylsulfonylmethane ($DMSO_2$) was used as a tracer (2μM) in each sample to allow for quantification of the methanol detected.

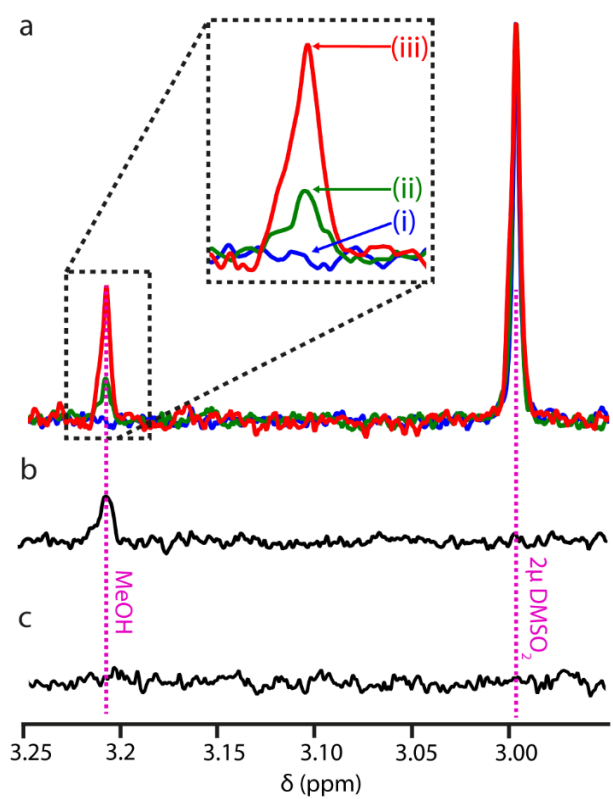

**Figure 2**. $^1$H NMR analysis for the methanol (MeOH) content of the ultrapure water solution after catalysis. (a) Spectra quantitatively comparing: (i) ultrapure water placed for 5 seconds in contact with the surface of HOPG, (ii) ultrapure water placed for 2 hours in contact with the surface of HOPG, and (iii) ultrapure water placed for 2 hours in contact with the surface of HOPG while applying a DC potential of +1 V to the HOPG. The peak at 3.21± 0.01 ppm is associated with the presence of methanol and the tracer peak (DMSO, peak just below 3 ppm) was used to adjust the relative magnitude of the curves. The determined concentrations of methanol are 0.116 ± 0.01 µM (i), 0 ± 0.01 µM (ii) and 0.295 ± 0.01 µM (iii). For comparison with samples (i-iii), spectra were also collected in a 1 µM solution of methanol (b) and in ultrapure water exposed to air for 2 hours (c). All samples were prepared in identical conditions (see methods for experimental details).

Sample (i) act as an immediate control for possible contaminants since no supramolecular structures could be observed by AFM on such short timescale. We chose a 5 seconds water-HOPG contact time as the shortest timescale in which the sample could be prepared. No detectable level of methanol is expected and this is indeed confirmed (Figure 2a (i)), demonstrating that the methanol present in the solution originates from catalytic activity and is not due to any form of external contamination.

Sample (ii) represents the shortest timescale typically necessary for the nucleation of supramolecular structures at the HOPG-water interfaces, as observed with AFM. A distinctive methanol peak is present (Figure 2a (ii)), confirming catalysis of water to methanol at the surface of HOPG immersed in water in ambient conditions. Comparing the area of the methanol peak with that of the $DMSO_2$ indicates a methanol concentration of 0.116 ± 0.01 µM in the solution. This is an underestimate since it excludes any methanol that remained at the HOPG surface after the solution was removed for analysis. Indeed, a significantly higher alcohol concentration is expected to remain at the interface with the hydrophobic HOPG surface where alcohol preferentially resides.[26,27] The present result shows that even at low concentrations, methanol can have a profound effect on the behavior of the interfacial liquid, as visible in Figure 1.

Sample (iii) serves two purposes. Firstly, the observed enhancement of the catalytic activity under an electrical potential further confirms the expected electrocatalytic activity of the HOPG both at atomic steps and edges where the existence of additional functional groups[19] may serve as oxidizing and reduction sites, similar to hydrophilic groups in GrO, and in the basal plane where fast electron transfer under applied fields is expected.[20] When a potential of +1V ± 0.01 V was applied for 2 hours, 0.295 ± 0.01 μM of methanol was detected (Figure 2a (iii)), more than twice the amount formed without the electric potential. Secondly, results from this sample suggest that the scanning AFM tip is likely to have an influence on the catalytic activity of the HOPG surface. Since catalysis can occur spontaneously at room temperature and is enhanced by the presence of an electric field, the silicon oxide AFM tip may also enhance the process. Silicon oxide tips such as those used in the present study develop a negative surface potential of typically -60 mV [28,29] when immersed in ultrapure water (pH of 5.8 in our experimental conditions), and the highly curved apex (<10 nm) can significantly enhance the resulting local electric field. Indeed, when water is placed in contact with the HOPG surface for several hours prior to imaging, ordered structures do not appear immediately at the start of imaging, but instead seem to be stimulated by the presence of the scanning tip.

Overall, the NMR results confirm spontaneous methanol catalysis. The process is enhanced by the presence of an electric field, potentially even as small as that induced by an immersed AFM tip at the interface with HOPG. Composite nanomaterials involving oxides are widely used in methanol catalysis, and could play a significant role here. We therefore decided to further investigate the effect of small voltages (comparable in magnitude to the surface potential of typical oxides) on the formation of interfacial structures. Since most of the alcohol produced resides at the interface with the HOPG, even small electric potentials could potentially have a significant impact.

Here we compared the evolution of samples exposed to 50 ± 1 mV or in open circuit for 24h. In both cases, it was necessary to de-wet the HOPG surface in order to transfer the pre-conditioned sample to the AFM chamber, thereby leaving a thin interfacial liquid layer containing the produced methanol. More ultrapure water is then added and the imaging starts within minutes. This procedure makes it difficult to rule out any disassembly/reassembly of the interfacial structures during the transfer, but the differences between the two samples are obvious nonetheless (Figure 3).

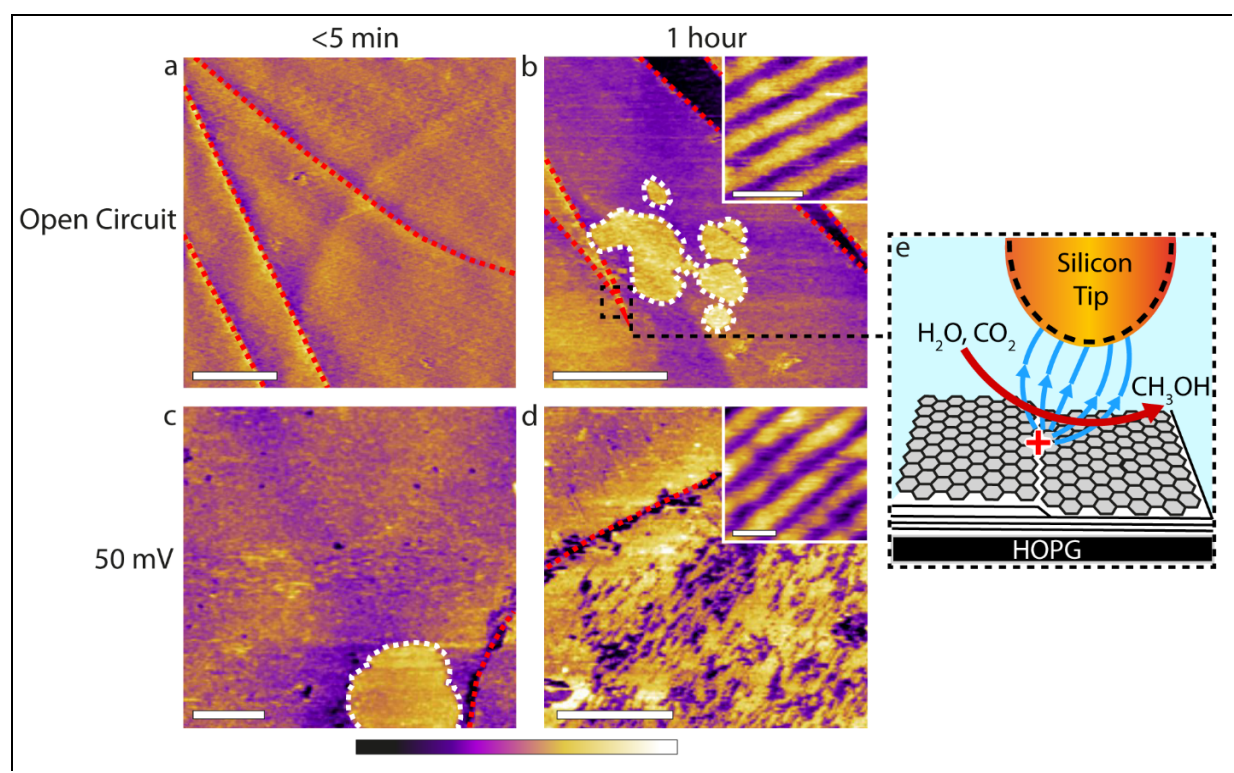

**Figure 3**. Influence of small electric potentials on the evolution of the methanol-water interfacial structures observed by AFM. (a) Image taken immediately after a sample that has been pre-conditioned for 24h in ultrapure water immediately after transferred into the AFM chamber, and (b) after 1 hour of imaging. Interfacial structures with row features (inset) outlined by the white dashes begin to appear near atomic steps (highlighted with red dashed lines). (c) Image taken immediately after a sample pre-conditioned for 24h in ultrapure water with an applied 50 mV DC potential was transferred into the AFM chamber. Some unstructured patches are already present on the surface. (d) After 1 hour of imaging the interfacial structures cover all the accessible area and also show the characteristic row

features (inset). (e) The hour long delay in nucleation between (a) and (b) despite the 24 hours of contact beforehand suggests the methanol production is electrocatalysed by the silicon AFM tip, which has a surface potential comparable to the electric field applied in (c) and (d). The white scale bars are 500 nm in the main images and 10 nm in the insets. The color scale bar represents a height variation of 1.5 nm in the main images and 1.2 nm in the insets.

Figure 3a shows that no features other than the characteristic graphite steps are initially visible in the absence of electrical pre-conditioning. After one hour of continuous imaging, small raised patches about 100 nm in diameter begin to nucleate near step-edges (Figure 3b). The patches exhibit supramolecular row patterns, comparable to those shown in Figure 1a. The assembly of these rows proceeds at a slow rate, here $0.14 \pm 0.1$ nm/s for the patch in Figure 3b, suggesting growth is limited by the rate of methanol catalysis. Indeed, previous work with water-methanol mixtures upwards of 5% volume concentration reported a growth of structures more than an order of magnitude faster than visible here.[22] Additionally, structures never fully cover the surface over the timescale of the experiment (several hours) with the largest patch observed exhibiting a diameter of ~1.2 μm.

In contrast, when a 50 mV external electric field is applied for 24h prior to imaging, unstructured patches form a monolayer on the HOPG surface and can immediately be observed (Figure 3c). The patches rapidly develop into ordered structures resembling those of the methanol-water monolayers. Significantly, within an hour of imaging the structures have almost completely covered the surface of the graphite (Figure 3d), at times forming multiple layers (see Figure S1 in the supporting information). While there are still nucleation sites near surface features, such as in Figure 3c, we also observe nucleation deep into the basal plane (see Figure S2 in the supporting information) indicating that the electrical pre-conditioning has caused electrocatalysis across all of the HOPG, instead of being limited to just surface singularities. These results all support the existence of a significantly larger quantity of methanol at the interface when compared to the sample prepared without any applied electric field. Furthermore, the hour long delay in nucleation between Figure 3a and Figure 3b despite the 24 hours of contact beforehand is consistent with the idea that while there is initially methanol present in the solution (as shown in the NMR data, Figure 2a (ii)), the local concentration is initially too low for nucleation, but the presence of the charged tip with a surface potential comparable to the electric field applied in Figure 3c and 3d, helps overcome this barrier and nucleation eventually takes place, Figure 3e.

AFM and NMR experiments consistently demonstrate catalysis of methanol at the surface of immersed graphite in ambient conditions. The mechanism allowing the methanol synthesized at the surface of graphite to be released in the bulk liquid is not immediately obvious from out data. Our results show that the water-methanol assembly is fully stable when directly in contact with the surface of graphite, but becomes progressively less stable as new layers form on top of the first layer. Additional layers could occasionally be observed (Figure S1) but the layers are only partially formed and exhibit many defects, suggesting a transition from stable assembly to bulk liquid. A single release mechanism is however unlikely and the methanol produced may be dispensed directly into the bulk liquid at catalytically active surface features where the water-methanol network is disrupted. Our results also suggest that the carbon source of the material converted into alcohol is primarily the carbon dioxide naturally dissolved in the ultrapure water. This is because the HOPG surface was not seen to evolve in time, ruling out any loss of material and carbon dioxide is the main source of volatile carbon in our experiment and is responsible for the slight acidity of ultrapure water in ambient conditions (pH 5.8). Catalytic conversion of carbon dioxide and water into methanol has been previously reported for GrO[18] where it is stimulated by white light (photo-catalysis). Here we did not observe any appreciable catalytic increase under illumination (see Figure S3 in the supporting information). However, experiments run in a sealed atmosphere for 2 hours (comparable in duration to sample (ii) in Figure 2) revealed no detectable concentration of methanol upon subsequent NMR analysis of the liquid (see Figure S4 in the supporting information), supporting the hypothesis of carbon dioxide as a reagent. We note that a degree of atmospheric contamination of the HOPG[25,30] is expected despite all the steps taken to minimize contamination. However, the consistent and reproducible trends observed indicate that possible contaminants do not dominate the results. Furthermore, the typical HOPG contaminants[25,30] are molecules far larger than methanol and so would not interfere with the AFM observations of its catalytic production.

The exact molecular mechanism underlying the catalytic process cannot by deduced from the present results alone, partly because the exact chemical details of the graphite are not know; surface groups at the edge plane[19] could significantly alter graphite's catalytic behavior by modifying the local electron accepting/donating abilities and inducing charge delocalization that would in turn impact chemisorption.[9] Our results suggest that the methanol production is caused by multiple factors, all involving the HOPG surface, that are difficult to disentangle. The AFM results provide consistent evidence of the AFM tip influencing the catalysis, likely through a tip field-effect[31] electrochemical reaction when in proximity to the HOPG surface (Figure 3). However, catalysis also occurs in the absence of an applied electric potential (Figure 2), through a mechanism dominated by step edges. Oxygen containing functional groups unique to multi-layered graphite could also be at play, inducing absorption via bonded and non-bonded interactions with the liquid molecules and serving as active sites.[32] In any case, the catalytic activity benefits from a positive polarization of the HOPG under an external electrical potential. This could be due to both a further enhancement of the chemical reaction at the step edges or to induced electron transfer occurring elsewhere. Indeed, recent studies have shown that doped graphene is able to reduce carbon dioxide in ambient conditions, when submitted to an electric potentials.[33]

When considering practical applications, the catalytic process reported here is far from optimized. It occurs slowly and is limited by the thermal energy available. In ambient conditions, we estimate a catalytic production rate of 4.6 mg $\times$ h$^{-1}$ $\times$ m$^{-2}$, far smaller than the best reported catalysis rates.[34] However, since singularities in the potential landscape such as atomic step-edges or the proximity of an AFM tip can significantly enhance the rate of catalysis, there is much scope for improvement. Additionally, the relatively low cost of graphite, its outstanding stability and the fact that material of technically lower quality (more defects) is catalytically more efficient should enable device geometries that maximize the catalytically active area without significant challenges.

From an electrochemical perspective, our results suggest that a graphite electrode immersed into an aqueous solution progressively develops an interfacial 'passivation' layer formed by a solid self-assembled layer of water and methanol molecules produced *in-situ*. While relatively easy to destroy, this layer reforms spontaneously and may significantly affect interfacial processes such as charge exchange[35] and molecular adsorption[36] as well as the catalysis of other molecules.[19] Interestingly, this result also suggests that local probe investigations of graphitic materials in aqueous solution may be prone to tip induced catalysis effects.

**Conclusion**

In conclusion, our combined AFM and NMR results consistently show that graphite is able to spontaneously catalyse methanol at room temperature. The amount of methanol produced is relatively modest and catalysis appears to occur almost exclusively at surface singularities on the graphite such as atomic steps and in the absence of any external input of energy. The underlying molecular mechanism remains unclear due uncertainties over the chemical composition of the graphite, but applying an external electrical potential across the interface considerably enhances the catalysis rate, even when due to the surface potential of nano-objects located near the interface. We believe our findings could have a significant impact on the development and understanding of novel carbon-based catalytic materials as well as devices highly sensitive to interfacial liquids.

EXPERIMENTAL SECTION

**Sample preparation.** All the solutions were prepared with ultrapure water (AnalaR NORMAPUR ISO 3696 Grade 3, VWR Chemicals, Leicestershire, UK) and HPLC-grade methanol with a purity of ≥99% (Sigma-Aldrich, Dorset, UK). In a typical experiment, a liquid droplet (~200 μL) of water was deposited on a freshly cleaved HOPG substrate (SPI supplies, West Chester, PA, USA) mounted on a stainless steel disk using silver paint (Ted Pella Inc, Redding, CA, USA). In all cases the HOPG was baked to >120 °C for 15 minutes to remove any contaminants[25] before depositing the droplet. The droplet was then left for a set period (5 seconds to 24 hours) inside a partially sealed glass container at room temperature (20 ± 1 °C). The container was thoroughly cleaned with ultrapure water beforehand and protected from the light throughout the incubation. The same procedure was used for the electric field experiments, except for a platinum wire (Sigma-Aldrich) is immersed in the droplet. The wire and HOPG sample were connected to a DC power supply (Aim-TTi, Cambridgeshire, UK) with a positive voltage applied to the HOPG with respect to the platinum.

**NMR.** After the determined incubation period the droplet was pipetted from the HOPG into a clean NMR tube. The solution was then diluted with deuterium oxide (purity 99.9%, Cambridge Isotope Laboratories, Inc., MA, USA) as needed and the $DMSO_2$ (Sigma-Aldrich, Dorset, UK) tracer added before conducting the measurement. The intense water signal was attenuated using the Robust-5 pulse sequence using a Varian (CA, USA) 600 MHz spectrometer equipped with an Agilent (CA, USA) probe able to deliver a maximum pulsed field gradient of 62 G cm$^{-1}$. Nine thousand two hundred and forty-eight scans were collected, each comprising 32728 complex data points and a spectral width of 10 kHz. The repetition time was 3.6 s, of which 1.6 s comprised the acquisition time. The W5 inter-pulse delay was set to 287 µs. Rectangular 1 ms pulsed field gradients were used with a strength of G =4.8 G cm$^{-1}$. The gradient stabilization delay was 1 ms. The error associated with estimating the quantity of methanol produced is dominated by the error in measuring the volume of liquid for NMR analysis, leading to an overall uncertainty on the concentrations of ±0.01 µM. The contribution to this error from the NMR measurement itself is negligible.

**AFM.** Imaging was conducted in amplitude modulation mode using a commercial Cypher ES AFM (Asylum Research, Santa Barbara, USA) equipped with temperature control and photothermal drive. The cantilevers (Arrow UHF-AUD, Nanoworld, Neuchatel, Switzerland) exhibit a nominal spring constant of ~1.95 nN/nm (from thermal spectrum calibration) and a resonance frequency of ~430 kHz in liquid. The cantilevers were cleaned by immersion in ultrapure water before imaging. All parts of the AFM in direct and indirect contact with the solution (cantilever holder, imaging chamber) where thoroughly cleaned with ultrapure water prior to imaging. After washing, the stage was heated to 105 ºC for 20 minutes to evaporate possible substances from previous experiments. In order to nucleate the structures[22] all the samples were imaged at 40 ºC, although all the preconditioning occurred at room temperature.

ASSOCIATED CONTENT

**Supporting Information**. The Supporting Information is available free of charge on the ACS Publications website at DOI: Additional AFM images showing multilayer growth (Fig S1), nucleation points far away from the step edges after the application of small electric fields (Fig S2), and control experiments testing the effect of light on the catalysis rate (Fig S3). Additional NMR spectra showing that no catalysis occurs in a sealed environment (Fig. S4).

AUTHOR INFORMATION

 **Corresponding Author**

*E-mail: kislon.voitchovsky@durham.ac.uk

**Author Contributions**

W. F., H. K. and K. V. designed all of the experiments. W. F conducted all the experiments except the NMR spectroscopy. W. F. prepared the samples for the NMR spectroscopy. J. A. A designed and conducted the NMR measurements and assisted with their analysis. All authors were involved in the writing of the paper and subsequent commenting.

ACKNOWLEDGMENT

This work was funded by the EPRSC through the Soft Matter and Functional Interfaces CDT (SOFI-CDT), Grant Reference No. EP/L015536/1. The authors would like to thank Dr. Alan Kenwright for his advice regarding the NMR spectroscopy and Prof. Bobby. G. Sumpter for useful discussions regarding the catalysis mechanism.

REFERENCES

(1) Wang, W.; Wang, S.; Ma, X.; Gong, J. Recent Advances in Catalytic Hydrogenation of Carbon Dioxide. *Chem. Soc. Rev.* **2011**, *40*, 3703–3727.

(2) Olah, G. A.; Goeppert, A.; Prakash, G. K. S. Beyond Oil and Gas: The Methanol Economy: Second Edition. *Beyond Oil Gas Methanol Econ. Second Ed.* **2009**, 1–334.

(3) Wu, C. T.; Yu, K. M. K.; Liao, F.; Young, N.; Nellist, P.; Dent, A.; Kroner, A.; Tsang, S. C. E. A Non-Syn-Gas Catalytic Route to Methanol Production. *Nat. Commun.* **2012**, *3*, 1050–1058.

(4) Huber, G. W.; Iborra, S.; Corma, A. Synthesis of Transportation Fuels from Biomass: Chemistry, Catalysts, and Engineering. *Chem. Rev.* **2006**, *106*, 4044–4098.

(5) Peppley, B. A.; Amphlett, J. C.; Kearns, L. M.; Mann, R. F. Methanol–Steam Reforming on Cu/ZnO/Al2O3. Part 1: The Reaction Network. *Appl. Catal. A Gen.* **1999**, *179*, 21–29.

(6) Dhakshinamoorthy, A.; Garcia, H. Catalysis by Metal Nanoparticles Embedded on Metal–Organic Frameworks. *Chem. Soc. Rev.* **2012**, *41*, 5262–5284.

(7) Graciani, J.; Mudiyanselage, K.; Xu, F.; Baber, A. E.; Evans, J.; D. Senanayake, S.;

Stacchiola, D. J.; Liu, P.; Hrbek, J.; Sanz, J. F.; Rodriguez, J. A. Highly Active Copper-Ceria and Copper-Ceria-Titania Catalysts for Methanol Synthesis from CO. *Science* **2014**, *345*, 546–550.

(8) Behrens, M.; Studt, F.; Kasatkin, I.; Kühl, S.; Hävecker, M.; Abild-Pedersen, F.; Zander, S.; Girgsdies, F.; Kurr, P.; Kniep, B.; Tovar, M.; Fischer, R. W.; Nørskov, J. K.; Schlögl, R. The Active Site of Methanol Synthesis over Cu/ZnO/Al2O3 Industrial Catalysts. *Science* **2012**, *759*, 893–897.

(9) Yu, D.; Nagelli, E.; Du, F.; Dai, L. Metal-Free Carbon Nanomaterials Become More Active than Metal Catalysts and Last Longer. *J. Phys. Chem. Lett.* **2010**, *1*, 2165–2173.

(10) Qu, L.; Liu, Y.; Baek, J. B.; Dai, L. Nitrogen-Doped Graphene as Efficient Metal-Free Electrocatalyst for Oxygen Reduction in Fuel Cells. *ACS Nano* **2010**, *4*, 1321–1326.

(11) Zhu, Y.; Murali, S.; Stoller, M. D.; Ganesh, K. J.; Cai, W.; Ferreira, P. J.; Pirkle, A.; Wallace, R. M.; Cychosz, K. A.; Thommes, M.; Su, D.; Statch, E. A.; Ruoff, R. S. Carbon-Based Supercapacitors Produced by Activation of Graphene. *Science* **2011**, *332*, 1537–1542.

(12) Chang, D. W.; Choi, H.-J.; Filer, A.; Baek, J.-B. Graphene in Photovoltaic Applications: Organic Photovoltaic Cells (OPVs) and Dye-Sensitized Solar Cells (DSSCs). *J. Mater. Chem. A* **2014**, *2*, 12136.

(13) Koenig, S. P.; Boddeti, N. G.; Dunn, M. L.; Bunch, J. S. Ultrastrong Adhesion of Graphene Membranes. *Nat. Nanotechnol.* **2011**, *6*, 543–546.

(14) Gil, A.; Colchero, J.; Luna, M.; Gómez-Herrero, J.; Baró, A. M. Adsorption of Water on

Solid Surfaces Studied by Scanning Force Microscopy. *Langmuir* **2000**, *16*, 5086–5092.

(15) Opitz, A.; Scherge, M.; Ahmed, S. I. U.; Schaefer, J. A. A Comparative Investigation of Thickness Measurements of Ultra-Thin Water Films by Scanning Probe Techniques. *J. Appl. Phys.* **2007**, *101*, 064310.

(16) Mi, B. Graphene Oxide Membranes for Ionic and Molecular Sieving. *Science.* **2014**, *343*, 740–742.

(17) Robinson, J. T.; Perkins, F. K.; Snow, E. S.; Wei, Z.; Sheehan, P. E. Reduced Graphene Oxide Molecular Sensors. *Nano Lett.* **2008**, *8*, 3137–3140.

(18) Hsu, H.; Shown, I.; Wei, H.; Chang, Y.; Du, H.; Lin, Y.; Tseng, C.; Wang, C.; Chen, L.; Lin, Y.; Chen, K. Graphene Oxide as a Promising Photocatalyst for $CO_2$ to Methanol Conversion. *Nanoscale* **2013**, *5*, 262–268.

(19) McCreery, R. L. Advanced Carbon Electrode Materials for Molecular Electrochemistry. *Chem. Rev.* **2008**, *108*, 2646–2687.

(20) Unwin, P. R.; Güell, A. G.; Zhang, G. Nanoscale Electrochemistry of Sp2 Carbon Materials: From Graphite and Graphene to Carbon Nanotubes. *Acc. Chem. Res.* **2016**, *49*, 2041–2048.

(21) Yang, D.; Zewail, A. H. Ordered Water Structure at Hydrophobic Graphite Interfaces Observed by 4D, Ultrafast Electron Crystallography. *PNAS* **2009**, *106*, 4112–4126.

(22) Voïtchovsky, K.; Giofrè, D.; Segura, J. J.; Stellacci, F.; Ceriotti, M. Thermally-Nucleated Self-Assembly of Water and Alcohol into Stable Structures at Hydrophobic Interfaces. *Nat. Commun.* **2016**, *7*, 13064.

(23) Aguilar, J. A.; Kenwright, S. J. Robust NMR Water Signal Suppression for Demanding Analytical Applications. *Analyst* **2016**, *141*, 236–242.

(24) Kondo, S.; Lutwyche, M.; Wada, Y. Nanofabrication of Layered Materials with the Scanning Tunneling Microscope. *Appl. Surf. Sci.* **1994**, *75*, 39–44.

(25) Martinez-Martin, D.; Longuinhos, R.; Izquierdo, J. G.; Marele, A.; Alexandre, S. S.; Jaafar, M.; Gómez-Rodríguez, J. M.; Bañares, L.; Soler, J. M.; Gomez-Herrero, J. Atmospheric Contaminants on Graphitic Surfaces. *Carbon* **2013**, *61*, 33–39.

(26) Sung, J.; Park, K.; Kim, D. Surfaces of Alcohol−Water Mixtures Studied by Sum-Frequency Generation Vibrational Spectroscopy. *J. Phys. Chem. B* **2005**, *109*, 18507–18514.

(27) Lundgren, M.; L. Allan, N.; Cosgrove, T. Wetting of Water and Water/Ethanol Droplets on a Non Polar Surface: A Molecular Dynamics Study. *Langmuir* **2002**, *18*, 10462–10466.

(28) Raider, S. I.; Flitsch, R.; Palmer, M. J. Oxide Growth on Etched Silicon in Air at Room Temperature. *J. Electrochem. Soc.* **1975**, *122*, 413–418.

(29) Shaw, A. M.; Hannon, T. E.; Li, F.; Zare, R. N. Adsorption of Crystal Violet to the Silica−Water Interface Monitored by Evanescent Wave Cavity Ring-Down Spectroscopy. *J. Phys. Chem. B* **2003**, *107*, 7070–7075.

(30) Li, Z.; Wang, Y.; Kozbial, A.; Shenoy, G.; Zhou, F.; McGinley, R.; Ireland, P.; Morganstein, B.; Kunkel, A.; Surwade, S. P.; Li, L.; Liu, H. Effect of Airborne Contaminants on the Wettability of Supported Graphene and Graphite. *Nat. Mater.* **2013**, *12*, 925–931.

(31) Huang, J.; Sumpter, B. G.; Meunier, V.; Yushin, G.; Portet, C.; Gogotsi, Y. Curvature Effects in Carbon Nanomaterials: Exohedral versus Endohedral Supercapacitors. *J. Mater. Res.* **2010**, *25*, 1525–1531.

(32) Sims, A.; Jeffers, M.; Talapatra, S.; Mondal, K.; Pokhrel, S.; Liang, L.; Zhang, X.; Elias, A. L.; Sumpter, B. G.; Meunier, V.; Terrones, M. Hydro-Deoxygenation of CO on Functionalized Carbon Nanotubes for Liquid Fuels Production. *Carbon* **2017**, *121*, 274–284.

(33) Song, Y.; Peng, R.; Hensley, D. K.; Bonnesen, P. V.; Liang, L.; Wu, Z.; Meyer, H. M.; Chi, M.; Ma, C.; Sumpter, B. G.; Rondinone, A. J. High-Selectivity Electrochemical Conversion of CO2 to Ethanol Using a Copper Nanoparticle/N-Doped Graphene Electrode. *ChemistrySelect* **2016**, *1*, 6055–6061.

(34) Studt, F.; Behrens, M.; Kunkes, E. L.; Thomas, N.; Zander, S.; Tarasov, A.; Schumann, J.; Frei, E.; Varley, J. B.; Abild-Pedersen, F.; Nørskov J. K.; Schlögl, R. The Mechanism of CO and CO2 Hydrogenation to Methanol over Cu-Based Catalysts. *ChemCatChem* **2015**, *7*, 1105–1111.

(35) Protsailo, L. V.; Fawcett, W. R. Studies of Electron Transfer through Self-Assembled Monolayers Using Impedance Spectroscopy. *Electrochim. Acta* **2000**, *45*, 3497–3505.

(36) Beckner, W.; He, Y.; Pfaendtner, J. Chain Flexibility in Self-Assembled Monolayers Affects Protein Adsorption and Surface Hydration: A Molecular Dynamics Study. *J. Phys. Chem. B* **2016**, *120*, 10423–10432.

**Graphic TOC**

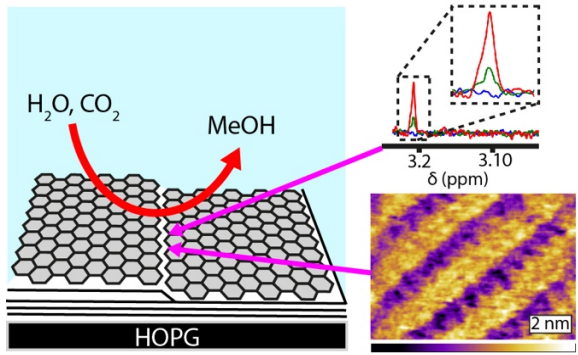

# Supporting Information

# *In-situ* molecular-level observation of methanol catalysis at the water-graphite interface

*William Foster[§], Juan A. Aguilar[†], Halim Kusumaatmaja[§] and Kislon Voïtchovsky[§]\**

[§]Durham University, Physics Department, Durham DH1 3LE

[†]Durham University, Chemistry Department, Durham DH1 3LE

\* Corresponding Author: kislon.voitchovsky@durham.ac.uk

**Content of the Supplementary Information (Fig S1-S4):**

– Fig. S1: AFM images of multilayer growth

– Fig. S2: AFM images of nucleation points far away from the step edges after the application of small electric fields

– Fig. S3: control AFM experiments testing the effect of light on the catalysis rate

– Fig. S4: NMR spectra showing that no catalysis occurs in a sealed environment.

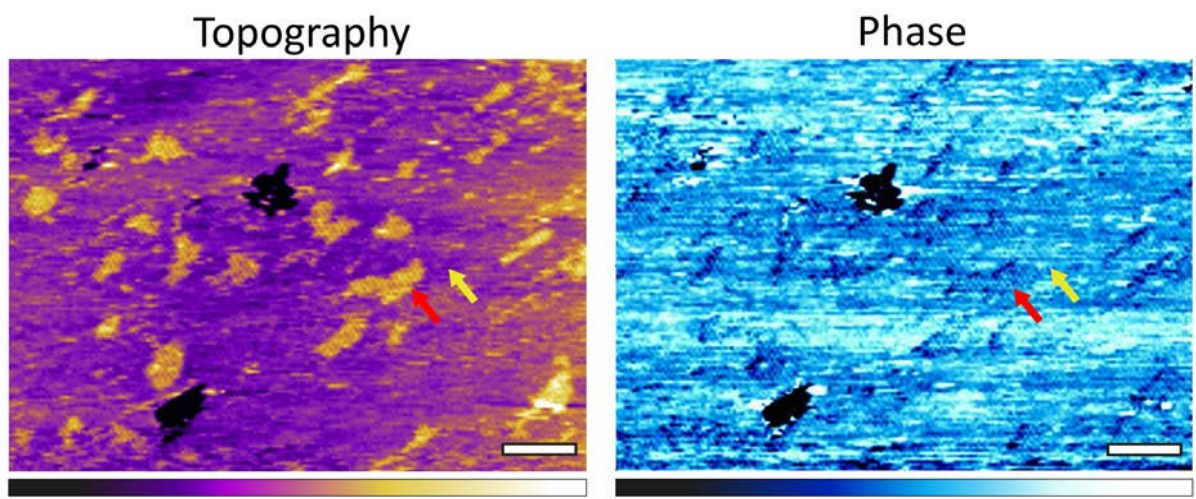

**Figure S1**. AFM imaging of the graphite surface preconditioned in a water droplet with a 50 mV electrical potential applied for 24 hours. The whole surface is covered with a structured monolayer (yellow arrow). A second layer can develop directly atop the first layer (red arrow). The nucleation of multiple layers suggests a higher concentration of methanol than when no electric field is applied where the structures exist only as monolayers. The white scale bars represent 100 nm. The purple scale bar represents a variation of 1.5 nm in the main images and blue scale bar represents a phase variation of 10°.

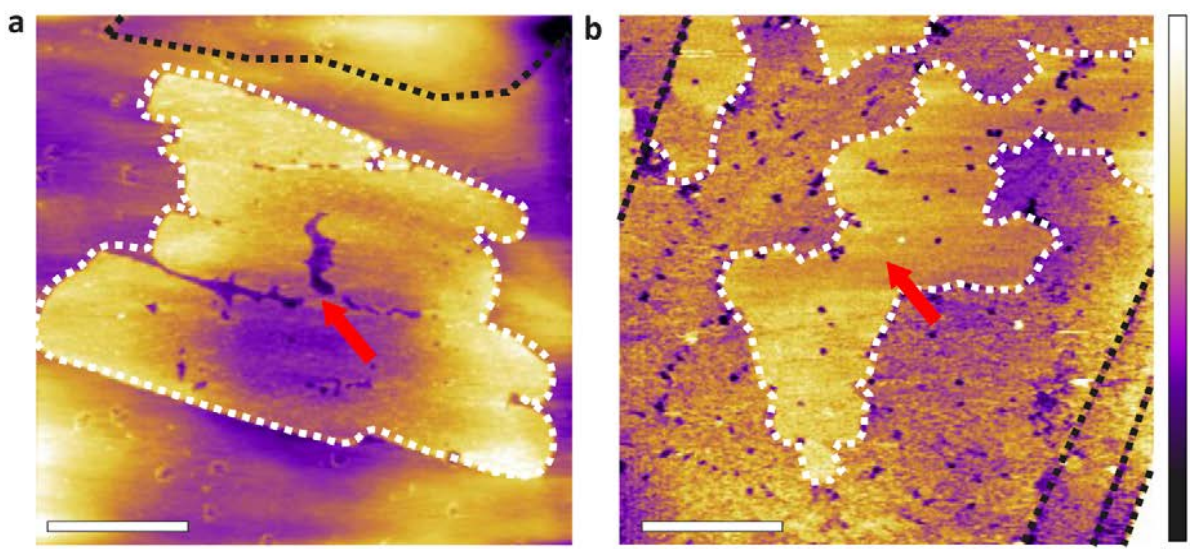

**Figure S2**. Topographic images of HOPG pre-conditioned for (a) 4h and (b) 24h in ultrapure water with an external 50mV DC electric field. The sample was immediately transferred to AFM chamber. In contrast to samples with no applied potential, nucleation sites can be found several microns away from step edges (red arrows), indicating the electrocatalytic activity is no longer limited to surface defects. The structures are outlined by white dashed line and the step edges highlighted with black dashed lines. The white scale bars represent 1 μm. The purple scale bar represents a variation of 4 nm.

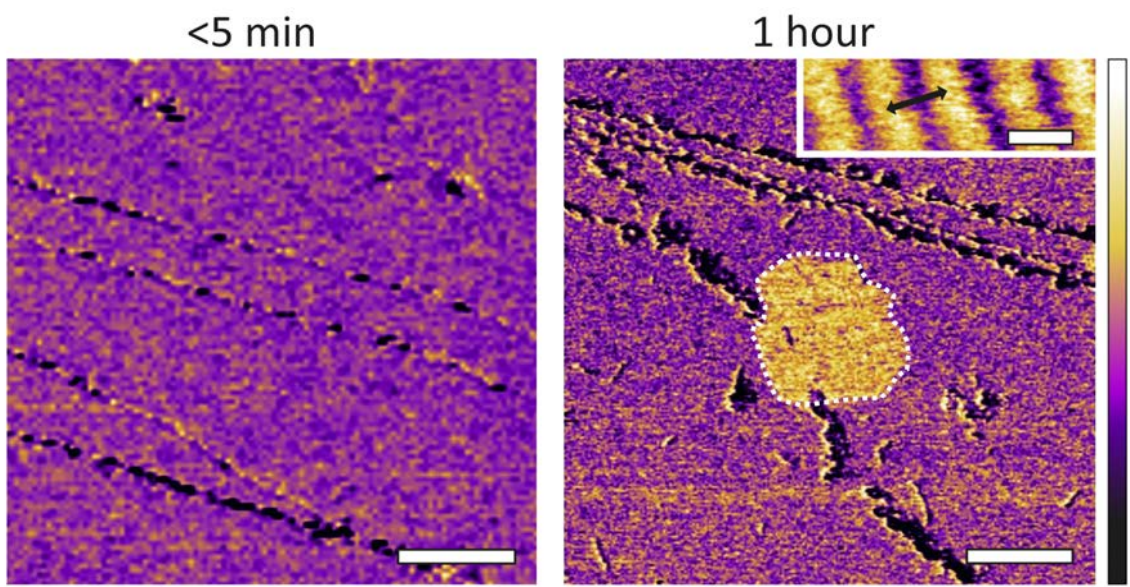

**Figure S3**. Amplitude modulation AFM phase images of HOPG pre-conditioned for 24h in ultrapure water with an external 60 W incandescent light bulb. The sample was immediately transferred to AFM chamber (left). The structures (outlined with the white dashed line) with a periodicity of 4.3 nm (black arrow) were only observed nucleating after an hour (right), consistent with the results in Figure 3 (a) and (b) indicating that no observable photo-catalysis of water to methanol has occurred. The inset is a topography image using the same color scale. The white scale bars represent 500 nm. The purple scale bar represents a variation of 5° in the main images and 1.5 nm in the inset.

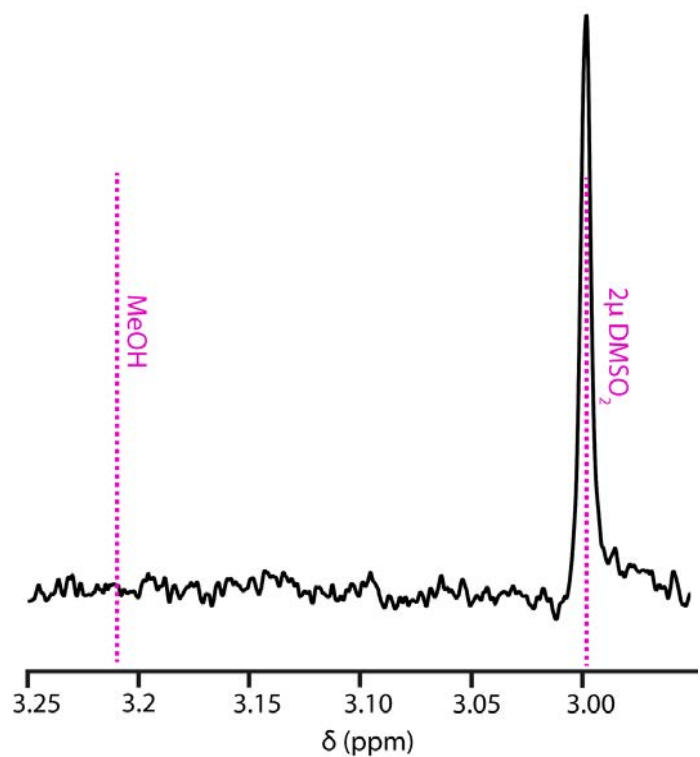

**Figure S4**. NMR spectrum for a droplet of ultrapure water left on HOPG for 2 hours in a sealed environment. The sample is doped with 2μM of Methylsulfonylmethane, consistent with the data in Figure 2. Here no methanol is detected, indicating the reaction most likely involves volatile organics that dissolve in the ultrapure water from the air.